\documentclass[preprint]{vgtc}                          

\graphicspath{{figures/}{pictures/}{images/}{./}} 
\usepackage{times}                     

\usepackage{tabu}    
\usepackage{enumitem}
\usepackage{booktabs}                  
\usepackage{lipsum}                    
\usepackage{mwe}                       

\usepackage{mathptmx}                  

\newcommand{\eg}{e.g.,}

\newcommand{\bpstart}[1]{\vspace{1mm} \noindent{\textbf{#1.}}}

\onlineid{0}

\vgtccategory{Research}

\vgtcinsertpkg

\title{Supporting Multimodal Data Interaction on Refreshable Tactile Displays: An Architecture to Combine Touch and Conversational AI}

\author{
    Samuel Reinders\thanks{e-mail:sam.reinders@monash.edu}\\ %
        \scriptsize Monash University %
    \and Munazza Zaib\thanks{e-mail:munazza.zaib@monash.edu}\\ %
        \scriptsize Monash University %
    \and Matthew Butler\thanks{e-mail:matthew.butler@monash.edu}\\ %
        \scriptsize Monash University %
    \and Bongshin Lee\thanks{e-mail:b.lee@yonsei.ac.kr}\\ %
        \scriptsize Yonsei University %
    \and Ingrid Zukerman\thanks{e-mail:ingrid.zukerman@monash.edu}\\ %
        \scriptsize Monash University %
    \and Lizhen Qu\thanks{e-mail:lizhen.qu@monash.edu}\\ %
        \scriptsize Monash University %
    \and Kim Marriott\thanks{e-mail:kim.marriott@monash.edu}\\ %
        \scriptsize Monash University %
}

\teaser{
  \centering
  \includegraphics[width=\linewidth]{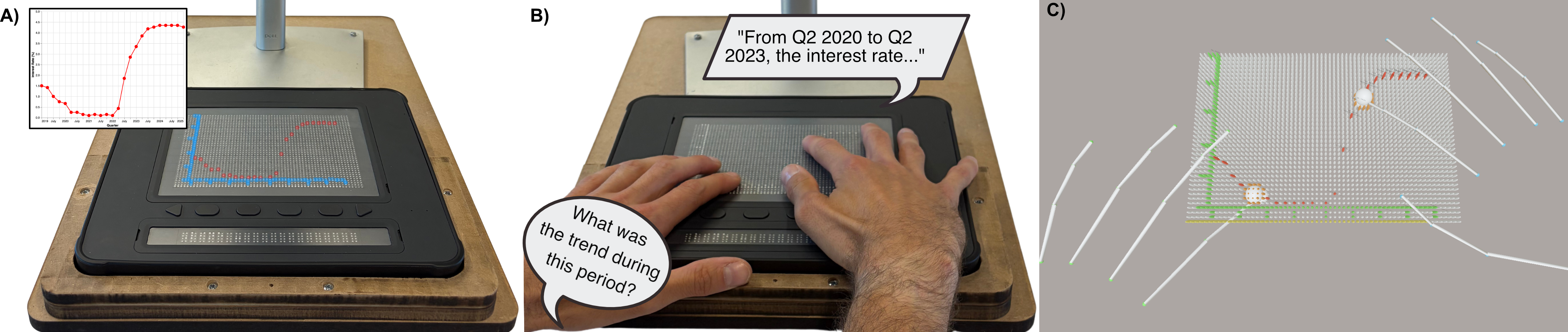}
  \caption{Our multimodal data interaction architecture combines touch and conversational AI on refreshable tactile displays for accessible data visualization. (A) A Vega-Lite chart rendered tactually on the Dot Pad's pin grid (data points and axis marks highlighted to improve legibility), with visual rendering shown as inset; (B) A user, having selected data points, asks a contextualized question; the agent responds with analysis grounded in the touched interval; (C) Unity view showing touch tracking.
  }
  \label{fig:teaser}
}

\abstract{Combining conversational AI with refreshable tactile displays (RTDs) offers significant potential for creating accessible data visualization for people who are blind or have low vision (BLV). To support researchers and developers building accessible data visualizations with RTDs, we present a multimodal data interaction architecture along with an open-source reference implementation. Our system is the first to combine touch input with a conversational agent on an RTD, enabling deictic queries that fuse touch context with spoken language, such as \textit{``what is the trend between these points?''} The architecture addresses key technical challenges, including touch sensing on RTDs, visual-to-tactile encoding, integrating touch context with conversational AI, and synchronizing multimodal output. Our contributions are twofold: (1)~a technical architecture integrating RTD hardware, external touch sensing, and conversational AI to enable multimodal data interaction; and (2)~an open-source reference implementation demonstrating its feasibility. This work provides a technical foundation to support future research in multimodal accessible data visualization.}


\begin{document}


\firstsection{Introduction}

\maketitle

Data visualizations are widely used in business, government, and everyday life. However, people who are blind or have low vision (BLV) face significant barriers in accessing data visualizations. Textual or spoken summaries alone are insufficient, as they do not provide for independent exploration or validation of findings~\cite{BANA2010guidelines}.

Refreshable tactile displays (RTDs), pin-based devices that render tactile graphics, offer promise for accessible data visualization. The availability of affordable RTDs, such as the Dot Pad~\cite{DotInc}, make tactile graphic delivery easier, faster, and more cost-effective than traditional production methods. At the same time, conversational AI agents, e.g., OpenAI's ChatGPT, have gained adoption, including among BLV users, for everyday information access and task completion. Combining RTDs with conversational AI addresses key limitations of each modality in isolation: conversational agents can provide immediate overviews and responses that cannot easily fit in Braille on RTDs' limited resolution displays, while RTDs provide spatial grounding of data and enable independent exploration and verification of information provided by conversational agents. 

Recent Wizard-of-Oz studies have shown that BLV users ask complex questions when exploring data visualizations~\cite{Kim2023} and prefer multimodal interaction combining touch exploration with conversational queries~\cite{Reinders2024}. No functional implementation exists that integrates touch input with conversational AI on an RTD for data visualization. This gap presents technical and architectural challenges, including: (C1)~existing RTDs lack touch sensing capabilities required for rich interaction; (C2)~visual chart encodings must be adapted to RTDs' tactile constraints; (C3)~touch context must be fused with speech queries to enable deictic interaction with the conversational AI (where users touch chart elements while asking questions that reference those touches); and (C4)~multimodal outputs across tactile, Braille labels, and audio must be synchronized.

The key contributions to the visualization and accessibility communities are:

\begin{enumerate}
    \item \textbf{A multimodal data interaction architecture} that addresses the four challenges (C1-C4) by integrating RTD hardware, external touch sensing, and conversational AI to enable accessible data visualization, including implementation details and design rationale.
    \item \textbf{An open-source reference implementation} that demonstrates and validates the architecture's feasibility.
\end{enumerate}

To our knowledge, this work presents the first system to combine touch input with a conversational AI agent on an RTD for accessible data visualization. Our implementation integrates the Dot Pad RTD with an Ultraleap hand tracker for continuous finger tracking (C1), renders Vega-Lite chart specifications as tactile graphics (C2), supports deictic queries that fuse touch context with spoken language through a conversational agent built on the GPT-4o model (C3), and synchronizes outputs across tactile, Braille, and auditory channels (C4). Three rounds of co-design workshops with three expert BLV co-designers refined the implementation, resulting in accurate responses and natural interactions. 

We release the implementation as open-source software to provide technical foundations for researchers and developers building accessible multimodal data visualization systems with RTDs.

\section{Background: Key Challenges and Related Work}
Data-driven decision making has become central across professions, yet the graphical nature of data visualizations creates accessibility barriers for BLV people. This has spurred research into accessible alternatives~\cite{lee2020reaching,kim2021accessible,marriott2021inclusive,dagstuhl2023inclusive,He2025}. Transcription guidelines recommend tactile graphics for accessible provision of maps, diagrams and graphs, rather than textual descriptions alone~\cite{BANA2010guidelines}, particularly where spatial relationships are important.

However, traditional tactile graphic production methods like using swell paper or embossing are not appropriate for interactive data exploration as each graphic requires substantial time and cost to produce. To address these bottlenecks, researchers have explored systematic generation approaches using declarative grammars like Tactile Vega-Lite~\cite{Chen2025}, multimodal systems that combine tactile graphics with audio descriptions~\cite{SharifEtAlCHI2022,alam2023seechart,Kim2023}, sonification (non-speech audio)~\cite{holloway2022infosonics,thompson2023chart,Ramoa2025,Chundury2024}, or refreshable tactile displays~\cite{elavsky2023data,holloway2024refreshable,Seo2024CHI} to support data analysis by BLV people.

In our prior work, we conducted a Wizard-of-Oz (WOz) study exploring how BLV users naturally interact with data visualizations with a RTD and conversational agent~\cite{Reinders2024,Reinders2024Poster}. Our findings revealed strong preference for multimodal interaction using touch and speech together, with users desiring multi-finger interaction for selecting intervals and comparing data points, and frequently making deictic references while exploring (e.g., touching points and asking \textit{``what happened here?''}). These WOz findings motivated the development of an architecture capable of supporting such interactions, but highlighted key technical challenges that we address in this work. This section reviews each challenge and related work.

\bpstart{C1: Touch Sensing on RTDs}
The first challenge in implementing rich multimodal interaction on RTDs is that existing devices lack adequate touch sensing capabilities. RTDs employ grids of electro-mechanically actuated pins~\cite{Yang2021} that raise and lower to form tactile patterns. Recent advances in RTD technology have produced more affordable devices including the Dot Pad~\cite{DotInc}, Monarch~\cite{APH}, and Graphiti~\cite{Orbit}.

However, touch sensing capabilities vary significantly across these devices. The Dot Pad does not support touch input. The Monarch uses an infrared touch sensor that supports only single-finger interaction with discrete touch modes, such as point-and-click and double-tap, and does not provide continuous positional tracking. The Graphiti, which incorporates an integrated touch-sensitive layer in the pin array, similarly supports only single finger, mode-dependent interaction.

These limitations conflict with many of the interaction behaviors observed in our WOz study, where BLV participants desired using multiple fingers simultaneously~\cite{Reinders2024}. Furthermore, supporting touch both for spatial exploration and as an input modality for selecting chart elements is challenging, as it requires a system to distinguish exploratory contact from intentional touch interactions. This gap between current RTD capabilities and user needs motivated the incorporation of external touch-sensing solutions.

\bpstart{C2: Visual-to-Tactile Encoding}
The second challenge is adapting visual chart encodings to the limited resolution and tactile constraints of RTDs. Current RTDs provide between 2,400 and 3,840 pins arranged in grids: the Dot Pad and Graphiti offer a 60$\times$40 grid, while the Monarch provides 96$\times$40. This constraint fundamentally limits the amount of data RTDs can present tactually.

Additionally, many visual encodings have no direct tactile equivalent on RTDs. Color and opacity, commonly used to distinguish data series in visual charts, have no direct translation on RTDs. This requires alternate encoding strategies, such as using different tactile patterns. Recent work addresses visual-to-tactile translation for printed tactile graphics. Tactile Vega-Lite~\cite{Chen2025} automatically generates tactile graphics from Vega-Lite specifications, adapting printed tactile constraints such as using appropriate tactile patterns and managing layout for embossing or swell paper printing. MAIDR~\cite{Seo2024ASSETS,Seo2024CHI} takes a different approach, using Braille unicode characters for sequential reading on single-line RTDs rather than creating two-dimensional spatial graphics.  

While these approaches address static or sequential reading, dense datasets introduce an additional encoding challenge: datasets with many observations might produce overlapping points that cannot be distinguished on the pin grid. Interactive exploration through zooming and panning could address this, but existing RTDs like the Monarch provide only geometric magnification (spacing pins farther apart) rather than semantic zooming (revealing detail on demand). This motivates a rendering pipeline capable of aggregation, interactive manipulation (zooming and panning), and semantic encoding that balances data fidelity with tactile legibility.

\bpstart{C3: Fusing Touch Context with Conversational AI}
The third challenge is fusing touch context with speech queries to enable deictic interaction. Participants in our WOz study made such deictic references while exploring data, touching specific points and asking questions like \textit{``what happened here?''} or making comparisons between touched points~\cite{Reinders2024}. Supporting this natural interaction pattern requires combining touch and speech inputs to ground conversational references in physical interactions. 

Advances in large language models (LLMs) have enabled conversational agents that can answer questions for data visualization~\cite{Zhao2025,kavaz2023}. Several systems have begun exploring LLM-based interfaces in BLV contexts, including: Vizability ~\cite{GorniakEtAl2024Vizibility} which allows users to query visual data trends using natural language, and MAIDR~\cite{Seo2024CHI,Seo2024ASSETS} which explored how BLV users leverage LLMs to interpret data visualizations. However, none of these systems support touch input, preventing deictic interaction where users touch chart elements while asking questions. The only work we know of that combines touch sensing and conversation is MapIO~\cite{ManzoniEtAl2024MapIO}, which supports deictic querying but focuses on tactile maps instead of data visualizations on RTDs. This gap between conversational chart exploration and deictic queries on RTDs motivated our architecture for combining touch context with natural language queries.

\bpstart{C4: Synchronized Multimodal Outputs}
The fourth challenge is synchronizing multimodal outputs to provide coordinated feedback and reduce confusion. Participants in our WOz study desired different outputs, including combination of speech, Braille labelling, and highlighting or pulsing different regions or specific points on the RTD. Without coordination across these modalities, users may receive conflicting or poorly-timed feedback that creates confusion about which chart elements or data points the response references. 

MAIDR~\cite{Seo2024CHI,Seo2024ASSETS} synchronizes positions across visual, tactile, textual, and auditory representations based on the user's current position, ensuring all modalities display the same data as users navigate. However, MAIDR does not support touch input or conversational responses with tactile feedback, and thus does not address the coordination challenges that arise from these interactions. 

Multi-line RTDs with touch input and conversational agents require both immediate coordination of multimodal outputs in response to user interactions (\eg\, simultaneously rendering tactile highlights, Braille labels, and audio when a user double-taps a data point) and temporal coordination during agent responses (\eg\, segmenting multi-sentence explanations into chunks synchronized with tactile highlights of referenced data points). This challenge motivated us to explore coordinating multimodal outputs to provide coordinated feedback across user interactions and agent responses. 

\vspace{-1mm}\section{Multimodal Data Interaction Architecture}

\begin{figure}[!htbp]
    \centering
    \includegraphics[width=1.0\columnwidth, trim=0 7.5mm 0 5mm, clip]{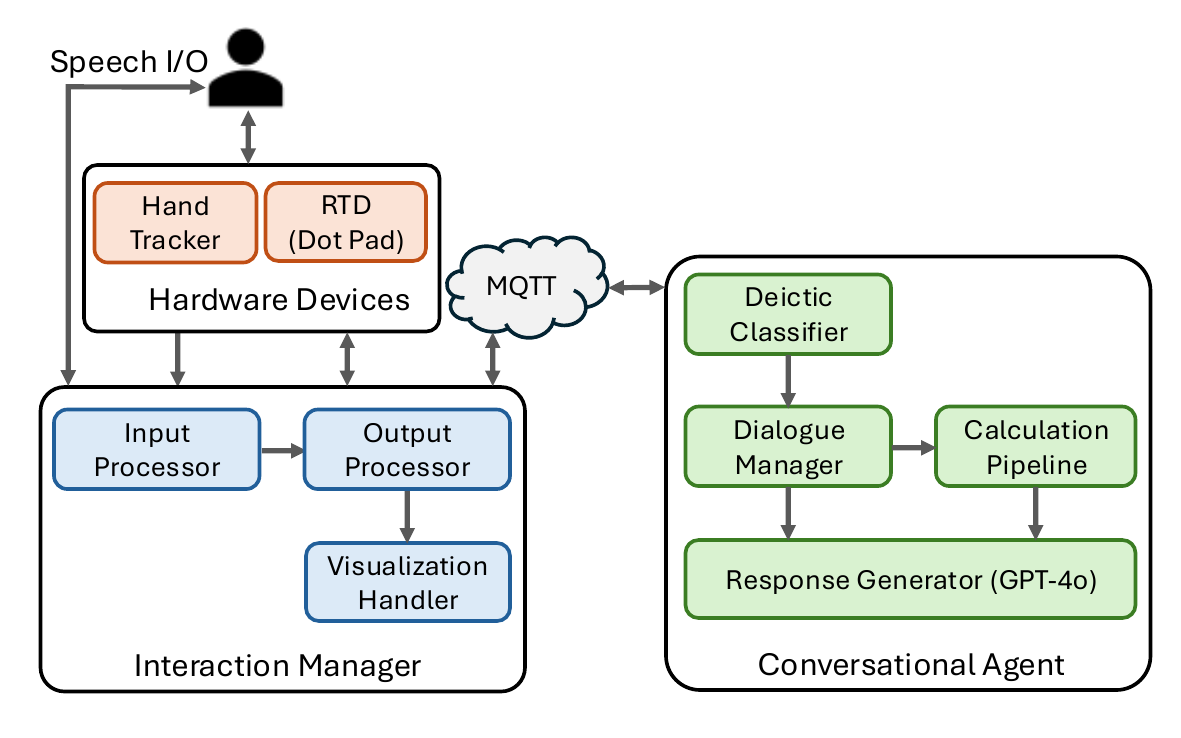}
    \vspace{-5mm}
    \caption{The multimodal data interaction architecture, consisting of the three interconnected components--Hardware Devices, Interaction Manager, and Conversational Agent--that coordinate via MQTT.}
    \label{fig:architecture}
    \vspace{-3mm}
\end{figure}

In this section, we present our multimodal data interaction architecture, which combines touch input with conversational AI on an RTD to support accessible data interaction. The architecture addresses the four main technical challenges described in Section 2. It comprises three core components (Figure~\ref{fig:architecture}): (1)~\textbf{Hardware Devices} capture touch and speech inputs and output tactile and auditory feedback; (2)~\textbf{Interaction Manager} processes inputs, renders charts, and coordinates communication; and (3)~\textbf{Conversational Agent} interprets queries that include speech and touch context and generates responses grounded in that context. 

Users may begin by exploring a chart rendered on the RTD through touch to identify regions or data points of interest, or by asking the conversational agent for an overview. Once familiar with a chart, users may perform deictic interactions that combine touch and speech. The following interaction scenario (Figure~\ref{fig:teaser}) illustrates how these components work together, with subsections detailing how each component processes the interaction:

\begin{enumerate}
    \item The user traces data points of the Interest Rates line chart, double-tapping 2020 Quarter 2 with their left index finger and 2023 Quarter 2 with their right. The system highlights the points, \textit{``2020 Quarter 2, interest 0.25\% ... 2023 Quarter 2, interest 3.85\%.''}
    \item They invoke the agent and ask: \textit{``What was the trend of the interest rate data during \textbf{this} period?''}
    \item The system responds, \textit{``From Q2 2020 to Q2 2023, interest rates declined from 0.25\% to 0.10\%, remained at 0.10\% for several quarters, and then increased steadily to 3.35\%.''}
\end{enumerate}
 
\subsection{Hardware Devices}
This component consists of hardware connected to a host computer: an RTD for tactile output and an external hand tracker.

\vspace{-1mm}\subsubsection{RTD Selection}
Our implementation uses the Dot Pad RTD because it is the most affordable multi-line tactile display on the market (under 5,000 USD), and provides a robust, publicly documented SDK with support for drawing operations, low-latency pin actuation, hardware button inputs, and API communication over Bluetooth LE or USB serial. We plan to add support for additional RTDs in the future.

\vspace{-1mm}\subsubsection{Touch Sensing}
\begin{figure}[!htbp]
    \centering
    \includegraphics[scale=0.25, trim=0 1mm 0 5mm, clip]{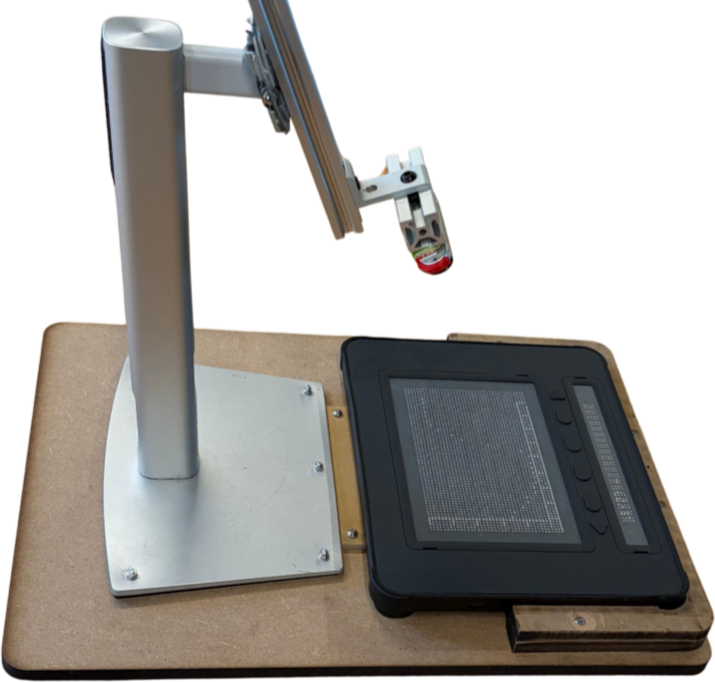}
    \caption{Custom mounting positioning the LMC 20 cm above the Dot Pad RTD at a 35$^\circ$ downward angle for stable tracking.}
    \label{fig:enclosure}
    \vspace{-3mm}
\end{figure}

As outlined in C1, both the Monarch and Graphiti support only single-finger interaction with discrete touch modes rather than continuous positional tracking. Our implementation augments the Dot Pad RTD with the Ultraleap Leap Motion Controller 2 (LMC), which provides continuous 3D tracking of multiple fingers simultaneously at 120 fps. We selected the LMC over alternatives (Google MediaPipe, Intel RealSense, and Microsoft Kinect) for its spatial accuracy, low latency, and its robust SDK plug-in that supports both the Unity\footnote{\url{https://github.com/ultraleap/UnityPlugin}} and Unreal development engines.
This decision accepts added hardware and setup complexity in exchange for multi-finger capability that no RTD currently provides. A custom mounting stand positions the LMC above the RTD to ensure stable alignment and consistent tracking coverage (Figure~\ref{fig:enclosure}).

\vspace{-1mm}\subsection{Interaction Manager}
The Interaction Manager comprises three subsystems---Input Processor, Visualization Handler, and Output Processor---which process inputs, render charts, and coordinate outputs. It uses the Unity engine because the LMC outputs 3D skeletal hand tracking data including depth information, requiring a 3D engine to determine when fingers are making contact or hovering over the RTD's surface. Additionally, our implementation uses Unity over alternatives like Unreal for its lower resource overhead and extensive SDK ecosystem, including native Ultraleap, Picovoice, and MQTT integrations. Communication with the Conversational Agent uses the Message Queuing Telemetry Transport (MQTT) protocol, chosen for its lightweight publisher-subscriber model that supports bidirectional asynchronous messaging.

The example interaction unfolds in three steps. In Step 1, the user double-taps two data points, triggering immediate tactile and auditory feedback from the Output Processor; in Step 2, the user asks a deictic query referencing those touched points, which the Input Processor captures and fuses with the touch context before sending to the Conversational Agent; and in Step 3, the Conversational Agent's response is routed back to the Output Processor for delivery as coordinated multimodal feedback. 

\subsubsection{Input Processor}
The input subsystem detects user interactions across speech, button presses, and touch gestures. It distinguishes exploratory touch from intentional gestures (C1) and captures touch context to enable deictic speech queries (C3).

\bpstart{Touch input and gesture recognition}
In Step 1 of the example interaction, when the user double-taps Q2 2020 with their left index finger and Q2 2023 with their right index finger, this component processes the LMC's skeletal hand data to distinguish between exploratory touch and intentional interaction (C1). The system maps 3D hand positions from the hand tracker onto the RTD's 2D pin-grid model aligned to the RTD's surface to detect touch contact. Only the index fingers are monitored, as they serve as natural pointers while avoiding false positives from unintentional contact during touch exploration. This design choice prioritizes accuracy; it limits gestures to two-finger interaction but still enables simultaneous selection for comparing data points and selecting intervals.

When contact occurs, a collision tracker uses a Gaussian-based inference model to determine the most likely target pin and chart element (data point or axis label). For the left index finger, it identifies Q2 2020 at grid coordinates [x1,y1] as the most likely target; for the right finger, it identifies Q2 2023 at [x2,y2]. Temporal analysis classifies gestures as taps or double taps. Double-tap gesture events for both touched points, along with their pin coordinates, data values (0.25\% and 3.85\%), and spatial probabilities, are cached to enable Step 2's deictic query resolution (C3).

\bpstart{Wake word and speech input} 
In Step 2, when the user begins to speak their request, the Picovoice Porcupine Wake Word engine monitors Unity's microphone audio for a custom wake word phrase, triggering speech recognition upon detection. We selected Porcupine for its low-latency local processing. Captured audio is streamed to Google Cloud Speech-to-Text, which returns the transcription \textit{``what was the trend of the interest rate data during this period?''} alongside word-level confidence scores. This transcript is then combined with the cached touch context from Step 1 (Q2 2020 and Q2 2023 coordinates and values) for deictic query interpretation by the Conversational Agent (C3).

\bpstart{RTD button input} 
The serial stream between the Dot Pad RTD and host computer is parsed to detect button presses from the RTD's six physical buttons (Left, F1, F2, F3, F4, and Right).

\vspace{-1mm}\subsubsection{Visualization Handler}
This subsystem addresses C2 by transforming Vega-Lite chart specifications into tactile representations optimized for the RTD's 60$\times$40 pin grid while supporting manipulation and semantic zooming. It currently handles quantitative, temporal, ordinal, and nominal data encodings across line charts, bar charts, scatterplots, and thematic maps. We chose Vega-Lite because it is a widely used declarative grammar and easily parsed and manipulated by LLMs.

\bpstart{Chart discovery and ingestion}
The chart discovery service scans for Vega-Lite JSON specifications, extracting metadata including chart name, field names, and column schema. A PNG preview is generated automatically using the Vega-Lite convert CLI utility\footnote{\url{https://github.com/vega/vl-convert}}. This metadata, along with the datasets and preview images (Base64), is compiled into a chart catalogue sent to the Conversational Agent for chart loading, multimodal chart understanding via the vision model, and data querying via the calculation pipeline.

\bpstart{Vega-Lite chart loader} 
After loading Vega-Lite chart specifications in response to the Conversational Agent's \textit{Load Chart} intent, the loader manages viewport state to handle datasets that exceed the RTD's resolution, tracking the current X-window, Y-window, and magnification level. Users can invoke zooming and panning operations via the Conversational Agent's \textit{Operations} intent or button shortcuts. Panning and geometric zooming change the viewport bounds, while semantic zooming triggers the data transformer to re-aggregate data at different resolutions (e.g., from weekly to daily aggregation of data), revealing additional detail rather than merely spacing pins further apart (C2). After changes to the viewport, the loader delegates rendering to the Vega-Lite-to-RTD renderer.

\bpstart{Vega-Lite data transformer} 
Support for Vega-Lite's data transformation operations: aggregation, calculation, filtering, and jitter, are implemented. The transformer applies transformations sequentially during chart load and re-applies them dynamically during zooming, panning, or layer switching. Multi-layer hierarchies allow datasets to be pre-computed at multiple resolutions (e.g., daily, weekly, monthly aggregations), allowing the chart loader to switch between aggregation levels during semantic zoom.

\bpstart{Vega-Lite-to-RTD renderer}
This component converts chart data into a tactile representation tailored for the RTD's pin grid, addressing C2. Unlike text-based approaches such as MAIDR that use Braille unicode characters, the renderer creates a two-dimensional spatial graphic. Semantic markers encode chart elements (background, axes, data points, scroll bars) to enable the Input Processor to identify touched elements. Chart elements are mapped to grid coordinates using linear interpolation, and a tactile zero-line is rendered when \textit{y}=0 falls within the current viewport.
    
\bpstart{Unity graph visualizer}
The system transforms rendered chart elements into structured data objects containing grid coordinates, values, and semantic types, enabling touch-based interaction (C1). The Output subsystem uses these to match finger positions to chart elements, capturing touched data for multimodal responses (C3).

\vspace{-1mm}\subsubsection{Output Processor}
The Output subsystem translates user interactions and agent responses into coordinated multimodal feedback across tactile, Braille, and auditory channels (C4). In Step 1 of the example interaction, it responds to the double-tap gestures with immediate tactile and auditory feedback. In Step 3, it processes and delivers the Conversational Agent's trend analysis response.

\bpstart{Button response handler} 
The Dot Pad RTD has six buttons (Left, Right, F1, F2, F3, F4) that support a three-state button model based on press duration and combinations. Quick taps ($< 200$ ms) on the Left and Right buttons page through Braille labelled content. Long holds ($\geq 500$ ms) trigger different functions: Left and Right navigate between data points; F1 activates push-to-talk voice input; F2 stops all outputs; F3 repeats audio; and F4 refreshes the display. Simultaneous button combinations (pressed within 100 ms of each other) enable panning and zooming operations.
    
\bpstart{Touch response handler}
To address C4, double-tap gestures trigger coordinated multimodal feedback across tactile, Braille, and auditory channels. In Step 1, when the user double taps the Q2 2020 and Q2 2023 data points, the handler coordinates tactile feedback by rendering a highlight pattern around both data points, and generates Braille labels and audio output via Google Cloud Text-to-Speech: \textit{``2020 Quarter 2, interest 0.25\% ... 2023 Quarter 2, interest 3.85\%.''}. Tactile highlights can be static or pulsed (animated raising/lowering) and persist temporarily or until dismissed, providing tactile confirmation of successful selection that users otherwise cannot obtain visually. As double-tapping is the main way of selecting elements, touch metadata for these two points, including data values, spatial probabilities, and pin coordinates, are cached for transmission to the Conversational Agent in Step 2's speech query, supporting deictic query resolution (C3).

\bpstart{Agent response handler} 
Responses from the Conversational Agent via MQTT are routed to appropriate output components. In Step 3, the handler processes the Conversational Agent's response. This response is synthesised using Google Cloud Text-to-Speech, and may include Braille labels, pin actuation commands (highlighting regions, drawing patterns, or pulsing specific points), or chart manipulation instructions (triggering the Visualization Handler). To maintain synchronization within multimodal responses, longer responses are segmented into navigable chunks by sentence boundaries. Each chunk is synchronized with corresponding tactile highlight patterns so users can track which data points are being described in the response. For example, a response like \textit{``In May 2021, interest rates increased to 0.25\%. In June they dropped to 0.1\%.''} would segment into two chunks, with the first and second highlighting the May and June data point, respectively, providing coordinated progression between verbal and tactile modalities (C4).

\bpstart{RTD communication}
Commands from the visualization and output subsystems are translated into binary packets for the Dot Pad. We chose USB serial communication over Bluetooth LE because it provides significantly higher data rates (115,200 baud $\approx$ 115 kbps vs 10 kbps) and lower latency for real-time updates.

\vspace{-1mm}
\subsection{Conversational Agent}
The Conversational Agent interprets natural language queries in the context of touch events and chart data, generating responses that can trigger multimodal feedback and visualization updates. In Step 2, the agent receives the query \textit{``What was the trend of the interest rate data during this period?''} along with the touch context from Step 1 indicating that the user touched Q2 2020 and Q2 2023. In Step 3, the agent generates a trend analysis spanning those points, which is sent back to the Interaction Manager's Output Processor for delivery. Grounding conversational references in physical touch interactions addresses C3. While the architecture supports the use of other LLMs, our implementation uses OpenAI's GPT-4o and LangChain. We chose GPT-4o for its multimodal capabilities and performance on data analysis tasks, while LangChain provides a tool-use framework enabling natural-language-to-code transformation and orchestration, better fitting our needs than retrieval-first frameworks such as LlamaIndex. The agent runs in a Jupyter Notebook and communicates with the Interaction Manager via MQTT.

\bpstart{Prompt design}
The agent's system prompt defines its role as a data visualization assistant for BLV users. The prompt design aligns with Gricean Maxims \cite{grice1975logic}: the \textit{Maxim of Manner} by requiring the agent to mention whether the user is touching a data point or an axis (C3), always begin with context before stating data, and give descriptive answers in no more than 40 words; the \textit{Maxim of Quantity} by including all values if the dataset has multiple max or min, and always including appropriate measurement units based on context; the \textit{Maxim of Quality} by not using the word ``approximately'' if the result is exact; and \textit{Maxim of Relation} by asking clarification questions if a question is ambiguous. 

We adopted zero-shot prompting over few-shot learning due to the multi-dataset nature of our architecture. Few-shot examples caused the agent to overfit to dataset-specific patterns, leading to incorrect responses when applied to other datasets. Zero-shot prompting enables generalization across chart types and data domains.

\bpstart{Deictic classifier}
This component determines whether a user's query contains deictic references (\eg\, \textit{``these points''} or \textit{``this bar''}). In Step 2, it detects the deictic reference \textit{``this period''} and fuses the touch context from Step 1 (Q2 2020 and Q2 2023) with the speech transcript when a confidence threshold (0.40) is met. The augmented query: \textit{``what was the trend... (touched: point\_A \{quarter=2020-Q2, interest=0.25\%\}; point\_B \{quarter=2023-Q2, interest=3.85\%\})''} enables the agent to interpret \textit{``this period''} as the range from Q2 2020 to Q2 2023 (C3).

\bpstart{Dialogue manager}
The manager classifies user queries into five intent categories: \textit{Load Chart}, \textit{Overview}, \textit{Image Analysis}, \textit{Operations}, and \textit{Data Explore}. For the trend analysis query in Step 2, the manager classifies it as \textit{Data Explore}. The first four are handled through dedicated procedures, while data exploration queries trigger a LangChain-based calculation pipeline. The manager maintains dialogue context to support multi-turn interactions.

\bpstart{Calculation pipeline}
Chart ingestion makes data available to the calculation pipeline as a pandas dataframe. For the example \textit{Data Explore} query in Step 2, GPT-4o transforms the query into executable code using LangChain: \texttt{df[(df['quarter'] >= '2020-Q2') \& (df['quarter'] <= '2023-Q2')]['interest'].describe()}. The PythonAstREPLTool executes the code, and results are returned to the LLM for natural language response generation (\eg\, identifying that rates declined to 0.10\%, remained stable, then increased to 3.35\%).

\bpstart{Response generator}
The generator produces natural language responses using GPT-4o. In Step 3, it generates: \textit{``From Q2 2020 to Q2 2023, the interest rate declined...''}. To support synchronized multimodal feedback (C4), responses include metadata identifying referenced chart elements or data points. The response is sent to the Interaction Manager, and when the agent mentions specific points, the Output Processor tactually emphasizes those referents to maintain grounding across modalities.

\vspace{-1mm}\section{Discussion \& Conclusion}

\noindent\textbf{Enabling Deictic Interactions on RTDs.}
We present the first system to successfully combine touch input with a conversational agent on an RTD for data visualization. This enables deictic queries where users touch chart elements while asking contextualized questions such as \textit{``what is the trend between these points?''}, grounding conversational analysis in spatial exploration. By addressing four key technical challenges (C1-C4), we demonstrate that rich multimodal interaction on RTDs is technically feasible. This system moves beyond our prior WOz findings~\cite{Reinders2024} that identified interaction preferences, providing a working technical foundation for realizing these interaction patterns.

\bpstart{Synchronization: From Invisible to Critical}
Implementing multimodal interactions revealed that synchronizing outputs across tactile, Braille, and auditory channels was critical for realizing the interaction patterns observed in our prior WOz study~\cite{Reinders2024}. While users naturally combine touch and speech when exploring data, coordination between these modalities is required in order to maintain coherence. For example, consider when a user double-taps a data point: a tactile highlight pattern, Braille label, and spoken output must be generated simultaneously. Agent responses introduce temporal complexity: longer explanations should be segmented into chunks synchronized with tactile highlights, letting users track which points are being described as the agent speaks. These synchronization challenges, both immediate and temporal, were invisible in our WOz setup, where a human wizard coordinated outputs, but became critical in our automated implementation. This highlights the gap between identifying interaction preferences through WOz studies and engineering systems that deliver them.

\bpstart{Vega-Lite as a Bridge to RTDs}
Our Vega-Lite-to-RTD renderer helps address the challenge of adapting visual chart specifications to RTDs' limited pin resolution, transforming Vega-Lite specifications into tactile representations with support for semantic zooming and data aggregation. Unlike text-based approaches such as MAIDR~\cite{Seo2024CHI} that use Braille unicode for sequential reading, our renderer creates two-dimensional spatial graphics for exploration through touch. We believe this renderer has value beyond our architecture: it could be used independently for generating static RTD graphics, and integrated into the Vega-Lite ecosystem similar to Tactile Vega-Lite~\cite{Chen2025} for printed tactile graphics. Future work should extend support for additional chart types and RTDs.

\bpstart{Complementary Modalities, Not Redundant Channels}
Our work enables tactile and conversational modalities to work together as complementary channels. Tactile exploration supports spatial understanding of chart structure, but struggles with precise value retrieval, while conversational queries excel at statistical calculations but lack spatial grounding. The combination of the two, supporting deictic queries, bridges this gap and can ground analysis within spatial contexts: BLV users can tactually explore to build spatial understanding, then ask contextualized questions about data points they've discovered. This design builds on prior work suggesting the complementary nature of modalities~\cite{Khalaila2025,Reinders2024} and provides technical foundations for investigating how orchestrating multiple channels together can leverage their respective strengths.

\bpstart{Architectural Foundations and Future Directions}
This work demonstrates that multimodal interaction combining touch and conversational AI on RTDs is technically achievable. Our architecture integrates RTD hardware, external touch sensing, and conversational AI. We release the architecture and open-source implementation\footnote{\url{https://github.com/accessible-data-vis/feelogue}} to provide technical foundations for the visualization and accessibility communities, advancing research in multimodal accessible data visualization. Our current implementation supports the Dot Pad RTD and four chart types (line charts, bar charts, scatterplots, and thematic maps) with a subset of Vega-Lite transformations. The architecture can be extended to support additional chart types. In addition, we plan to conduct a user study focused on the usability of the system, perceptions regarding the agent's responses, and accuracy of touch input. We hope our work inspires further research in multimodal accessible data visualization, enabling BLVs to perform independent data analysis. The current setup requires a custom mount and external sensor which limit portability; we also hope to motivate RTD hardware manufacturers to integrate multi-touch capabilities into RTDs, enabling interaction techniques such as pinch-to-zoom and multi-finger gestures. 

\acknowledgments{We gratefully acknowledge support from the Australian Research Council through the Discovery Projects funding scheme (DP220101221). This work was also supported in part by the Yonsei University Research Fund (2025-22-0099).}

\bibliographystyle{abbrv-doi}

\bibliography{template}
\end{document}